\documentclass[prc,preprint,preprintnumbers,amsmath,amssymb]{revtex4-1}

\usepackage{bm,curves}
\usepackage{epsfig}
\usepackage[colorlinks=false,linkcolor=blue]{hyperref} 
\usepackage{color}
\usepackage{dcolumn}
\usepackage{amsmath}
\usepackage{natbib}

\DeclareGraphicsExtensions{.ps}

\newcommand{\lan}{\langle}
\newcommand{\ran}{\rangle}
\newcommand{\EA}{{\it et al.}}

\begin{document}

\preprint{NT@UW-16-01}

\title{The Role of Nucleon Strangeness in Supernova Explosions}

\author{T. J. Hobbs$^1$}
\email{tjhobbs@uw.edu}

\author{Mary Alberg$^{1,2}$, Gerald A. Miller$^1$}
\affiliation{
        $^1$\mbox{Department of Physics,
         University of Washington, Seattle, Washington 98195, USA} \\
        $^2$Department of Physics,
         Seattle University,
         Seattle, Washington 98122, USA
}

\date{\today}

\begin{abstract}
Recent hydrodynamical simulations of core-collapse supernova (CCSN) evolution have
highlighted the importance of a thorough control over microscopic
physics responsible for such internal processes as neutrino heating. In
particular, it has been suggested that modifications to the
neutrino-nucleon elastic cross section can potentially play a crucial
role in producing successful CCSN explosions. One possible source
of such corrections can be found in a nonzero value for the nucleon's
strange helicity content $\Delta s$. In the present analysis, however, we show
that theoretical and experimental progress over the past decade has suggested
a comparatively small magnitude for $\Delta s$, such that its sole effect
is not sufficient to provide the physics leading to CCSN explosions.
\end{abstract}
\maketitle

%
%
%
Supernova (SN) explosions have long been recognized as the means of populating
the galaxies with heavier elements \cite{Burbidge:1957vc}. Despite much progress
\cite{Janka:2012wk}, however, the physics responsible for these events remains murky.

For core-collapse supernovae (CCSNe) specifically, the eventual explosion depends upon 
the evolution of the bounce shock produced by the implosion of a massive
star's Fe core to proto-neutron star \cite{Janka:2012wk, Burrows:2012ew}. Problematically,
dissipative effects from nuclear dissociation/neutrino emission
stall the advancing shock in numerical simulations, raising the
question of what mechanism re-energizes the outwardly moving front.
Since the early work of \cite{Bethe:1984ux}, the conventional explanation is
that delayed neutrinos reheat the post-shock region of the CCSN, reinvigorating
the shock front's advance. This picture has placed a considerable
premium upon controlling the various physics effects \cite{Horowitz:2001xf} that go into the
neutrino-nucleon interactions inherent to the ``delayed-neutrino mechanism.''

This is especially true in the vicinity of the gain radius ---
the locus where cooling from neutrino emission is in approximate equilibrium with
the corresponding heating due to neutrino absorption \cite{Burrows:2012ew}. As such, even
incremental modifications to the neutrino-nucleon cross section such as might
follow from corrections to the nucleon's flavor structure could
alter the explosive evolution of CCSNe. Effects of this kind might then in turn
play a decisive role in numerical simulations that successfully produce CCSN
explosions.

The importance of such considerations was brought to the fore by the recent
publication of Melson \EA~\cite{Melson:2015spa}, which studied the impact of a
large magnitude nucleon strange helicity $\Delta s=-0.2$ in a simulated
CCSN originating from a $20\ M_\odot$ progenitor star. As we
shall describe, computing with $\Delta s=-0.2$ induces a $\sim$15$\%$ reduction
in the strength of the neutrino's axial coupling to the neutron, leading to
the diminution of neutron opacities to neutrinos relative to a calculation
using $\Delta s=0$, for which no explosive behavior was obtained. Using
$\Delta s=-0.2$ resulted in a calculation of a CCSN explosion $\sim$300~$ms$
after the shock bounce in Ref.~\cite{Melson:2015spa}. In practice, this large
amplitude for $\Delta s$ served as a proxy for $\mathcal{O}(\sim\!\! 15\%)$
corrections to neutrino-nucleon opacities that might arise from various
effects; here for definiteness, however, we study the specific plausibility
of such large values for $\Delta s$, given present knowledge of nucleon
structure.
%
%
%
%
The relevant physics is the neutrino-nucleon total elastic cross section,
to lowest order in the weak coupling \cite{Horowitz:2001xf}:
\begin{equation}
\sigma_i(\epsilon)\,=\,{2G^2_F\epsilon^2\over3\pi}\,\Big(c^2_{Vi}+5c^2_{Ai}\Big)\,.
\label{eq:cross}
\end{equation}
Here, $\epsilon$ is the energy of the incident
neutrino, and $i$ an isospin label.
Electroweak physics specifies the values of the nucleon couplings of the weak vector
and axial-vector (or, ``axial'' below) currents $c_{Vi}$ and $c_{Ai}$, which for $SU(2)$ are known
to be \cite{Agashe:2014kda}
\begin{align}
c_{Vp}\ &=\ {1\over2}-2\sin^2\theta_W\ ,\hspace*{1.0cm}c_{Vn}\ =\ -{1\over2}\ ;\nonumber\\
c_{Ap}\ &=\ +{g_A\over2}\ ,\hspace*{2.55cm}c_{An}\ =\ -{g_A\over2}\ ,
\label{eq:SU2_couplings}
\end{align}
where $g_A\approx1.26$ and $\sin^2\theta_W\approx0.2325$.
These effective couplings are the main input to the neutrino-nucleon physics,
and they potentially receive corrections from other flavor sectors
--- especially strange.

Differences of spin-conserving, axial current matrix elements can ultimately be related to the quark
helicity content of the nucleon. That is, by considering nucleonic matrix elements of the
weak axial current (here, in the light flavor $SU(2)$ sector with isospin label
$i=p,\ n$)
\begin{equation}
\label{eq:weak_cur}
\lan P',\lambda';i\,\big|J^\mu_A\big|P,\lambda;i\ran\
=\,\bar{u}_i(P',\lambda')\,\Big(\,\gamma^\mu\gamma_5\,{\tau_3\over2}\,G_A(Q^2)
+\dots\,\Big)\,u_i(P,\lambda)\,,
\end{equation}
the axial form factor may be accessed through the appropriate helicity difference of
the neutral current ``$+$'' components as defined on the light-front (for reviews
of the light-front formalism, see \cite{Brodsky:1997de,Bakker:2013cea,Miller:2000kv}):
\begin{align}
\pm{1\over2}G_A(Q^2)\ =\ {1\over4P^+}\Big( \lan P',\lambda'=\,\uparrow;i\,\big|J^+_A\big|P,\lambda=\,\uparrow;i\ran\,
-\,\lan P',\lambda'=\,\downarrow;i\,\big|J^+_A\big|P,\lambda=\,\downarrow;i\ran\Big)\,,
\label{eq:GAdef}
\end{align}
in which $P,\ P'$ are the $4$-momenta of the initial/final state nucleon, and
$\lambda,\lambda'=\,\uparrow,\downarrow$ their associated helicity designations; the $\pm$ above
corresponds to $i=p,\,n$. As usual, the form factor is scaled in terms of the elastic
momentum-transfer $Q^2=-(P'-P)^2$.

By merit of the parity-oddness of the $\gamma^+\gamma_5$ operator that weights
$G_A(Q^2)$ in Eq.~(\ref{eq:weak_cur}) for $\mu=+$, the axial form factor at $Q^2=0$ is directly
related to the quark-level spin content of the nucleon as well as the effective neutrino-nucleon coupling
\cite{Horowitz:2001xf} ---
\begin{equation} 
c_{Ai}\,=\,{1\over2}\,\Big(\pm\! G_A(0)\,-\,G^{s\bar{s}}_A(0)\Big)=\,{\pm g_A-\Delta s\over2}\,,
\label{eq:ax_coupling}
\end{equation} 
in the absence of electroweak radiative corrections \cite{Armstrong:2012bi}. The fact that it is the negative
of $\Delta s$ which enters Eq.~(\ref{eq:ax_coupling}) may be interpreted as resulting from the isoscalar nature of the
strange axial form factor $G^{s\bar{s}}_A$ relative to the isovector $G_A$; the former quantity
is the analog of the $SU(2)$ axial form factor defined in
Eqs.~(\ref{eq:weak_cur}) and~(\ref{eq:GAdef}), but evaluated in a basis that couples the axial
current to nucleon strangeness. In Eq.~(\ref{eq:ax_coupling}) we have also made the explicit
identification $G^{s\bar{s}}_A(0)=\Delta s$.
From the result of Eq.~(\ref{eq:ax_coupling}) it is straightforward to infer the qualitative impact
of nonzero strange helicity in the nucleon. Analyses generally prefer $\Delta s \le 0$, such that
\begin{align}
\Big(c^{(\Delta s\le0)}_{Ap}\Big)^2\, \ge\,\Big(c^{(\Delta s=0)}_{Ap}\Big)^2\,,
\hspace*{2cm}
\Big(c^{(\Delta s\le0)}_{An}\Big)^2\, \le\,\Big(c^{(\Delta s=0)}_{An}\Big)^2\,,
\end{align}
which leads to the respective enhancement and suppression of the $\nu-p$ and $\nu-n$ cross
sections given by Eq.~(\ref{eq:cross}) for $\Delta s\le0$ relative to the zero strange quark
helicity scenario.

Notably, the question of the $Q^2\sim0$ behavior of $G^{s\bar{s}}_A(Q^2)$ is of central
importance to the proton spin problem \cite{Thomas:2008bd}, wherein the total $1/2$ helicity of
the proton is decomposed in the usual fashion among quark helicities, orbital angular momenta,
and gluon total angular momentum according to
\begin{align}
{1\over2}\,=\,{1\over2}\,\Delta\Sigma_q+L_q+J_g\,,
\hspace*{2cm}
\Delta\Sigma_q=\Delta u+\Delta d+\Delta s\,.
\end{align}
In the expression above, $\Delta s$ contains contributions from both $s$ and $\bar{s}$, and we
emphasize that $\Delta q\equiv\lan\Delta q+\Delta\bar{q}\ran$, such that $\Delta\Sigma_q/2$
yields the quark helicity contribution to the proton spin.

Thus, given its fundamental importance, $\Delta s$ has been modeled/computed in various ways,
and been the focus of multiple experimental efforts, as discussed below.
%
%
%
%
Assuming it to be the sole correction to the $\nu-N$ cross section of Eq.~(\ref{eq:cross})
as in Ref.~\cite{Melson:2015spa}, the magnitude of $\Delta s$ used to successfully produce an
exploding CCSN is at odds with the latest hadron structure calculations and measurements.
Citing the lower bound of the 2002 analysis by Horowitz~\cite{Horowitz:2001xf}, which obtained
$\Delta s=-0.1\pm0.1$, the hydrodynamical simulations of Ref.~\cite{Melson:2015spa} were performed
at the extreme lower value $\Delta s=-0.2$.
The result $\Delta s=-0.1\pm0.1$ found in Ref.~\cite{Horowitz:2001xf}
was obtained from the 1987 $\nu-p$, $\bar{\nu}-p$ elastic scattering measurements at BNL \cite{Ahrens:1986xe}.
It must be stressed that this original analysis did not endeavor to determine $\Delta s$
specifically, but rather simply included a correction term $\eta$, fitted to the
neutrino-proton data, which was intended to parametrize potential modifications to the weak axial
current regardless of origin. For this and other reasons pointed out by Kaplan and Manohar~\cite{Kaplan:1988ku},
the results of the analysis in Ref.~\cite{Ahrens:1986xe} should be taken with caution regarding
the nucleon's strange helicity.
The value $\Delta s=-0.2$ used in Ref.~\cite{Melson:2015spa} lies well beyond the
range of the analyses/measurements that have emerged in the decade
following Ref.~\cite{Horowitz:2001xf}. These determinations of $\Delta s$ come from the
complementary directions of theory and phenomenological analyses of relevant data, as
we describe below.

%
%
%
Among newer sources of experimental information, global analyses of quark helicity
distributions constrained by data (mainly, spin-polarized parity-violating deeply
inelastic scattering [DIS] of electrons on nucleons)
constrain the magnitude/sign of the total strange helicity, permitting a somewhat
larger contribution to the proton spin. For example, the parton distribution function (PDF)
analyses summarized in Ref.~\cite{Diehl:2007uc} include those of Gl\"uck, Reya, Stratmann, and
Vogelsang \cite{Gluck:2000dy}; Bl\"umlein and B\"ottcher \cite{Bluemlein:2002be};
Leader, Sidorov, and Stamenov \cite{Leader:2005ci}; the Asymmetry Analysis Collaboration
\cite{Hirai:2006sr}; and de Florian, Navarro, and Sassot \cite{deFlorian:2005mw}. These
yield moderate values for the total strange helicity, resulting in the average value
$\Delta s = -0.120 \pm 0.021$. These analyses proceed by assuming a
parametric form for the Bjorken $x$ dependence of the quarks' helicity
distributions $\Delta s(x,Q^2_0)$ and $\Delta\bar{s}(x,Q^2_0)$ at the boundary $Q^2_0$
of a numerical QCD evolution scheme. The total strange helicity asymmetry is then
\begin{align}
\Delta s\,=\,(s^{\raisebox{-4pt}{+}}-s^{\raisebox{-3pt}{-}})+(\bar{s}^{\raisebox{-4pt}{+}}-\bar{s}^{\raisebox{-3pt}{-}})\
=\,\int_0^1\,dx\,\Big[\Delta s(x)+\Delta\bar{s}(x)\Big]\,,
\label{eq:del_s}
\end{align}
and $s^\pm$ represent spin-dependent distributions for
strange quarks with helicities parallel/antiparallel to that of their parent nucleon.
A persistent limitation of this approach is the lack of experimental
constraints on the parametric form at $Q^2_0$, as well as at the distribution endpoints,
particularly $x\sim0$ where the necessary extrapolation to evaluate
Eq.~(\ref{eq:del_s}) is especially fraught.

Since the PDF-based analyses described in \cite{Diehl:2007uc}, there have been other
direct measurements of $\Delta s$, occasionally yielding somewhat smaller results.
Of these, we briefly describe a representative sample.

Direct observations of spin asymmetries have enabled more precise extractions of
$\Delta s$ in the past several years. For instance,
in 2007 HERMES measured the total strange helicity at intermediate $Q^2=10$ GeV$^2$ \cite{Airapetian:2006vy},
obtaining $\Delta s=-0.09\pm0.02$. Given its form in Eq.~(\ref{eq:del_s}),
the $Q^2$ dependence of $\Delta s$ is governed by flavor-singlet QCD evolution, and the corresponding moment
at the perturbative starting scale $Q^2_0$ may therefore differ slightly. Moreover, these
determinations were highly sensitive to the fragmentation function (FF) parametrization, which is
inherently nonperturbative.
This source of systematic uncertainty inspired additional dedicated efforts to extract $\Delta s$
at COMPASS \cite{Alekseev:2009ac, Alekseev:2010ub}. Of these, Ref.~\cite{Alekseev:2010ub} reported several values of
$\Delta s$ at $Q^2=3$ GeV$^2$ using different extrapolation schemes. Employing a direct extrapolation
to $x=0,1$ as well as the DSSV parametrization, respectively, COMPASS obtained
\begin{align}
\Delta s &=-0.02\pm0.02\hspace{1cm} {\rm (extrapolation)}\,,\nonumber\\
&=-0.1\pm0.02 \hspace{1.18cm} {\rm (DSSV)}\ .
\end{align}

\begin{figure}
\vspace*{0.1cm}
\hspace*{0.05cm}
\centering
\includegraphics[scale=0.45]{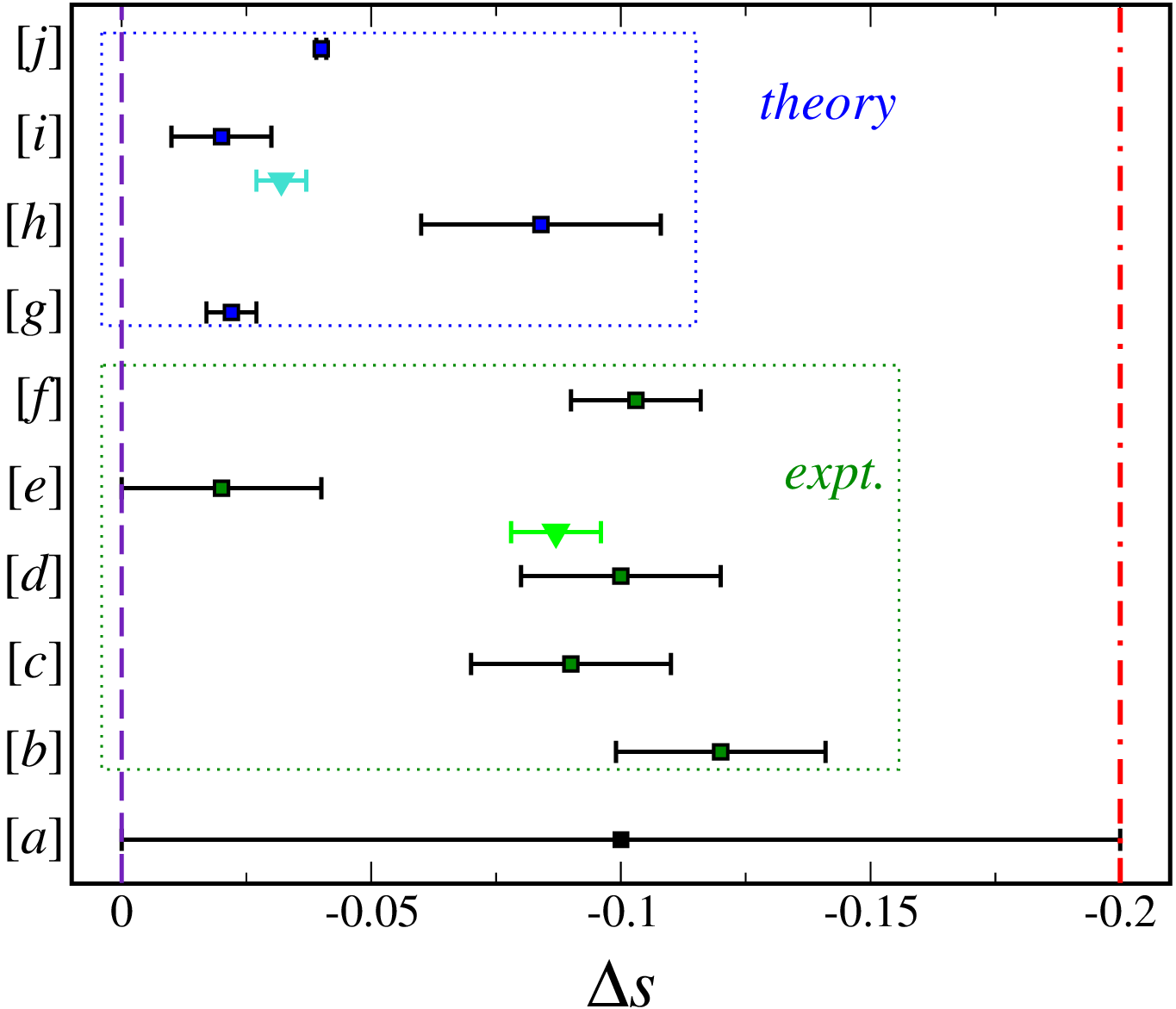}
\caption{A summary of $\Delta s$ determinations in this analysis.
Experimental/phenomenological findings subsequent to Ref.~\cite{Horowitz:2001xf} (solid,
black square, [{\it a}]) are contained in the lower rectangle. These
include the averaged results in \cite{Diehl:2007uc} [{\it b}], \cite{Airapetian:2006vy} [{\it c}], the
extrapolation and DSSV-based methods of \cite{Alekseev:2010ub} [{\it d} and {\it e},
respectively], and the average of \cite{Leader:2014uua} and \cite{Sato:2016tuz} [{\it f}].
The lower inverted triangle represents the mean of these measurements: $\Delta s=-0.087\pm 0.009$.
Modern theoretical calculations are given in the top rectangle, including the averaged
small-$\Delta s$ lattice calculations
\cite{QCDSF:2011aa,Babich:2010at,Engelhardt:2012gd,Abdel-Rehim:2013wlz,Chambers:2015bka}
[{\it g}], the AWI-based computation of \cite{Gong:2015iir} [{\it h}], \cite{Bass:2009ed}
[{\it i}], and \cite{Hobbs:2014lea} [{\it j}]. The upper inverted triangle depicts the theory
average: $\Delta s=-0.032\pm0.005$, while the vertical dot-dashed line represents
the value used in Ref.~\cite{Melson:2015spa}.
}
\label{fig:bar}
\end{figure}

Finally, we also mention the result of a recent re-analysis of the spin-dependent strange PDFs
$\Delta s(x)$ and $\Delta\bar{s}(x)$
\cite{Leader:2014uua}, made in light of new CLAS data on the proton spin structure function $g^p_1(x,Q^2)$
\cite{Prok:2014ltt}. Computing the strange portion of the nucleon spin from PDFs constrained by
these data, the result $\Delta s=-0.106\pm 0.023$ followed \cite{Gong:2015iir}. Meanwhile, another
comprehensive global analysis of these data was reported by Sato \EA~\cite{Sato:2016tuz}, in this case
making use of a novel, Monte Carlo-based fitting scheme. This calculation arrived at
a comparable central value and range for the moment of the twist-2 strange spin-PDF,
$\Delta s=-0.10\pm 0.01$. We report the average of these two results ($\Delta s=-0.103 \pm 0.013$)
as well as the other experimental information in Fig.~\ref{fig:bar}, noting that the above
determinations correspond to somewhat different $Q^2$, which introduces some modest uncertainty.

%

%
%

While the precision of experimental data presently allows values of the
strange helicity as large as $\Delta s\sim-0.1$, various theoretical
calculations are considerably more stringent, several examples of which we highlight below.
In lattice gauge theory, contributions from strange quarks to nucleonic matrix elements
inherently arise from disconnected diagrams whose evaluation is vastly more
computationally expensive. Technical improvements, however, as well as the exponentiating
availability of the necessary computational resources have now rendered such calculations
more feasible.

Thus, in recent years $\Delta s$ has become amenable to calculation using lattice
techniques as carried out by, {\it e.g.}, the QCDSF collaboration \cite{QCDSF:2011aa}. This group
obtains $\Delta s=-0.020\pm0.010\,(stat.)\pm0.004\,(sys.)$.
Meanwhile, a separate lattice collaboration \cite{Babich:2010at} has
preliminarily calculated $\Delta s$ (albeit {\it without} renormalization effects),
also finding a small, negative result: $\Delta s=-0.019\pm0.011$.
As opposed to these $N_f=2$ analyses, however, work described in Ref.~\cite{Engelhardt:2012gd} employed the
2+1-flavor gauge configurations of the MILC Collaboration \cite{Toussaint:2009pz}, yielding the
renormalized value $\Delta s=-0.031\pm0.017$ at the physical pion mass. Ref.~\cite{Abdel-Rehim:2013wlz}
found the slightly larger value $\Delta s=-0.0227\pm0.0034$ in
a setting wherein the axial charges were comprehensively evaluated. Similarly, the analysis of
Ref.~\cite{Chambers:2015bka} obtained $\Delta s=-0.018\pm0.006$ using a numerical scheme motivated
by the Feynman-Hellmann theorem adapted to the lattice, although this study and its predecessors [with the
exception of \cite{Engelhardt:2012gd}] were performed with unphysically large pion masses. These various
lattice computations exhibit a general concordance, and for them we find an average of
$\Delta s=-0.022\pm0.005$.

Unlike these efforts, the recent calculation of the $\chi$QCD Collaboration
\cite{Gong:2015iir} explicitly enforced the anomalous Ward identity (AWI) through
a normalization factor $\kappa_A$ weighting local axial-vector currents on the finite lattice.
Following chiral extrapolation to the physical pion mass, this produced a comparatively large
value relative to the other lattice works, $\Delta s=-0.084\pm0.024$.

Other model calculations are also possible under the auspices of various theoretical
frameworks. For example, the Cloudy Bag Model calculation of Ref.~\cite{Bass:2009ed}, which
followed the older calculation presented in \cite{Tsushima:1988xv}, incorporated strange
quarks into nucleon structure by including kaon-hyperon fluctuations in the extended
meson cloud of the proton. Doing so, the flavor-$SU(3)$ axial charges could be
evaluated in the context of the MIT Bag Model, resulting in the values
\begin{align}
g^{(0)}_A\,&=\,\Delta u+\Delta d+\Delta s\,=\,0.37\pm0.02\,,\nonumber\\
g^{(8)}_A\,&=\,\Delta u+\Delta d-2\Delta s\,=\,0.42\pm0.02\,,
\end{align}
from which one may derive the similarly small \linebreak
$\Delta s=(g^{(0)}_A-g^{(8)}_A)/3=-0.02\pm0.01$. This method thus yields
extremely close agreement with the aforementioned small-$\Delta s$ lattice calculations,
aside from \cite{Gong:2015iir}.

In yet another model-based analysis, Hobbs \EA~\cite{Hobbs:2014lea} investigated the strange content of the
proton using the framework of light-front wave functions (LFWFs). Truncating the nucleon Fock expansion
at a two-body quark/scalar tetraquark state containing $s$ and $\bar{s}$, Ref.~\cite{Hobbs:2014lea} obtained a universal
wave function for the strange content of the nucleon with the ability to interpolate
between elastic form factor measurements and DIS structure functions in the strange
sector: 
\begin{align}
\big|\Psi^\lambda_P(P^+,{\bf P_\perp})\ran\,=\,{1\over16\pi^3}\sum_{q=s,\bar{s}}\int{dx\ d^2{\bf k_\perp}\over\sqrt{x(1-x)}}\
\psi^\lambda_{q\lambda_q}(x,{\bf k_\perp})\,
\big|q;xP^+,x{\bf P_\perp}+{\bf k_\perp}\ran\,,
\label{eq:Fock-2}
\end{align}
wherein the internal light-front $4$-momentum of the strange quark interacting with the weak axial current
is $(k^+,{\bf k}_\perp,k^-)=(xP^+,{\bf k}_\perp,k^-)$ in a frame with zero transverse momentum for
the nucleon, ${\bf P}_\perp={\bf 0}_\perp$.
By then constraining the helicity wave functions $\psi^\lambda_{q\lambda_q}(x,{\bf k_\perp})$ to
unpolarized DIS measurements, Ref.~\cite{Hobbs:2014lea} found novel bounds on the elastic observables
$\mu_s$ and $\rho_s$. In addition, with a phenomenological determination of the strange quark
wave function of the nucleon, the strange sector matrix element analogous to Eq.~(\ref{eq:GAdef})
could be computed, which led to the result
\begin{equation}
-0.041\le\Delta s\le-0.039\,,
\end{equation}
in line with the small strange helicity magnitudes of the above-mentioned theoretical calculations. We
summarize this sampling of theoretical calculations of $\Delta s$ along with the above-mentioned
experimental information in Fig.~\ref{fig:bar}.

\begin{figure}
\vspace*{0.5cm}
\centering
\includegraphics[scale=0.6]{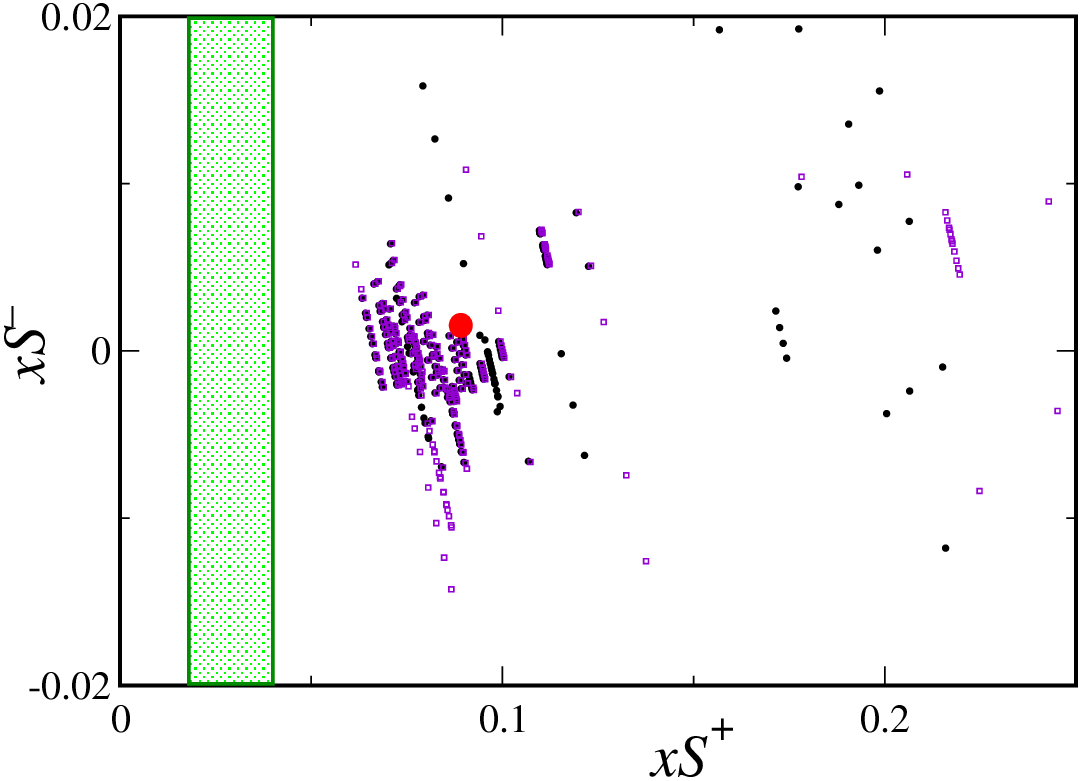}
\caption{Parameter scans for the 
\cite{Hobbs:2014lea} model consistent with $\Delta s=-0.20\pm0.01$.
Results using wave functions with Gaussian (dots)
and dipole-like (squares) momentum dependence are displayed. Parameter
combinations satisfying $\Delta s\sim-0.2$ require large
wave function normalizations, generating the lower bound $xS^+\ge0.062$.
This substantially overshoots the range (vertical band) for
$xS^+$ determined by CTEQ \cite{Lai:2007dq}. The large disk near $(xS^+,\,xS^-)=(0.09,\,0.002)$
represents the model used for the strange PDFs plotted in Fig.~\ref{fig:HERMES}. }
\label{fig:scan}
\end{figure}
\begin{figure}
\hspace*{-0.3cm}
\centering
\includegraphics[scale=0.45]{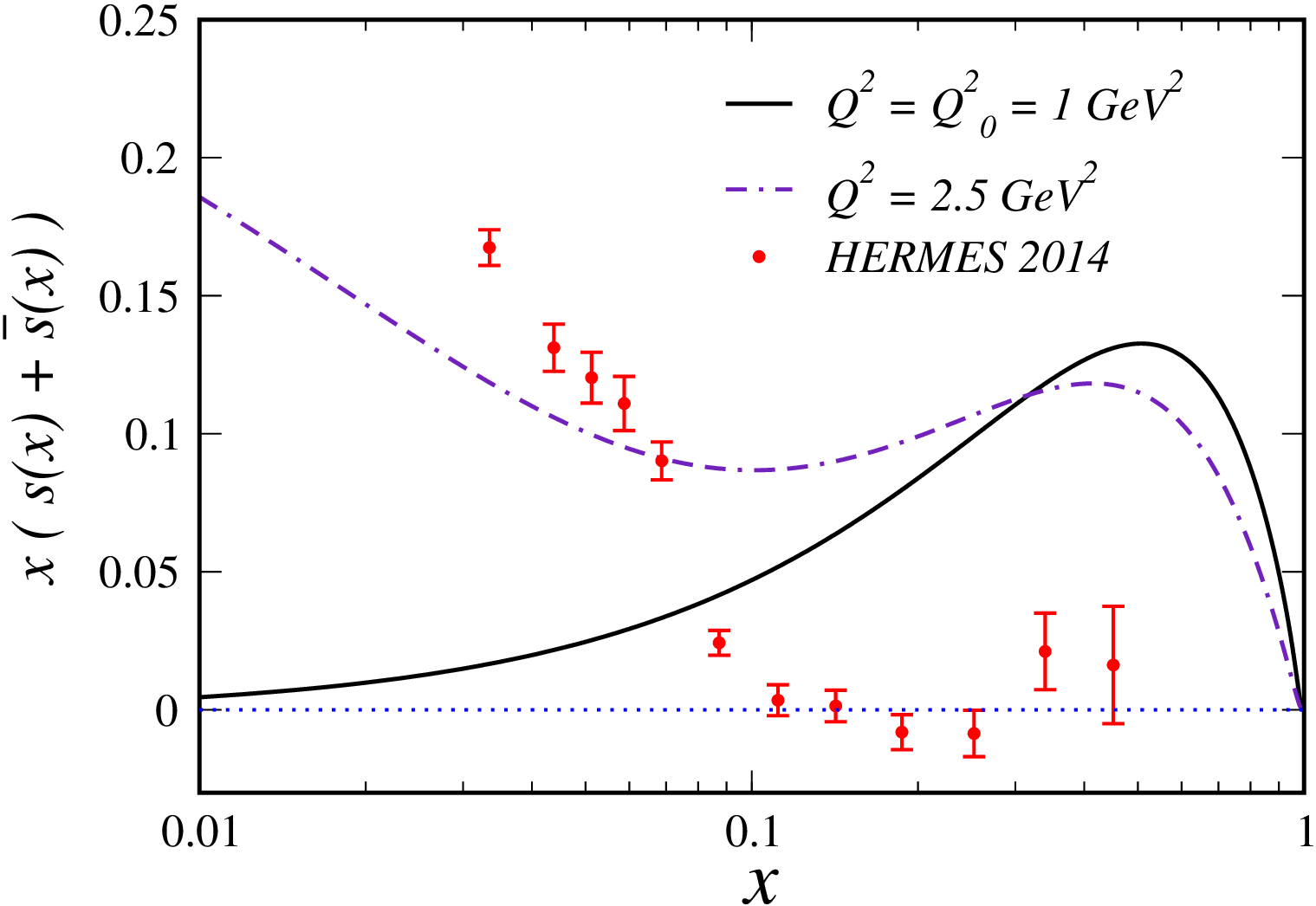}
\caption{
$x[s(x)+\bar{s}(x)]$ consistent with $\Delta s=-0.2$ according to the model of
\cite{Hobbs:2014lea}. The model parameters leading to the above correspond to
the large dot at $xS^+\sim0.09$ in Fig.~\ref{fig:scan}. The model has been evolved
from $Q^2_0 = 1$ GeV$^2$ (solid line) to $Q^2=2.5$~GeV$^2$ (dot-dashed line) for
comparison with HERMES \cite{Airapetian:2013zaw} (red circles). A model constrained
to $|\Delta s|$ comparable to that of Ref.~\cite{Melson:2015spa} thus hugely over-estimates the
strange PDF --- especially for $x\gtrsim0.1$.
}
\label{fig:HERMES}
\end{figure}

In fact, by making use of the same numerical approach employed by \cite{Hobbs:2014lea} to explore the
parametric LFWF model space, we find that a strange helicity asymmetry as large as
$\Delta s=-0.2$ as in Ref.~\cite{Melson:2015spa} would require magnitudes for nucleon strangeness
in the DIS sector as ruinously large as $xS^+\sim0.1$, where we define
\begin{equation}
xS^\pm\,\equiv\,\int^1_0\,dx\,x\Big[s(x)\pm\bar{s}(x)\Big]\,.
\end{equation}
Such a sizable value is highly excluded by global analyses of the world's high
energy data \cite{Lai:2007dq,Martin:2009iq}. We obtain this by performing systematic scans
of the parameter space of the model in Ref.~\cite{Hobbs:2014lea} and admitting for
consideration only those combinations consistent within $5\%$ of
$\Delta s=-0.2$. The resulting locus of points representing these
specific model parameter combinations may then be plotted in a
two-dimensional space spanned by $xS^-$ {\it vs.} $xS^+$ as shown
in Fig.~\ref{fig:scan}. We note that, while the smallest value of $xS^+$
tolerated by the model for $\Delta s=-0.2$ occurs for $xS^+=0.062$,
the mean of the full set corresponds to $xS^+=0.1$. In either case
this far exceeds the range determined by global analyses of DIS data such as a recent
dedicated CTEQ strangeness study \cite{Lai:2007dq}, which found the $90\%$ limit
$0.018\le\,xS^+\le\,0.04$ indicated by the green band in Fig.~\ref{fig:scan}.
Similarly, a simple estimate using the fitted distributions of MSTW at
the QCD evolution starting scale $Q^2_0=1$ GeV$^2$ leads to the comparable
$68\%$ range $0.017\lesssim\,xS^+\lesssim0.033$ \cite{Martin:2009iq}.
As a further demonstration, we consider a particular model taken from the
space plotted in Fig.~\ref{fig:scan}, and use it to compute the unpolarized
quark PDF combination $x[s(x)+\bar{s}(x)]$. This has been measured by
various experimental collaborations, including HERMES \cite{Airapetian:2013zaw},
and we compare a model prediction associated with $\Delta s=-0.2$ with
these data at $Q^2=2.5$ GeV$^2$ in Fig.~\ref{fig:HERMES}. According to
expectation given the large values of $xS^+$ shown in Fig.~\ref{fig:scan}
for $\Delta s=-0.2$, the evolved model plotted in Fig.~\ref{fig:HERMES}
as the dot-dashed curve egregiously overhangs the experimental data
obtained via kaon production in semi-inclusive DIS.
A general and robustly motivated light-front model is therefore difficult
to reconcile with large strange quark helicities without simultaneously predicting
implausibly large values for the total strange momentum $xS^+$ and $x$-dependent
distribution $x[s(x)+\bar{s}(x)]$. This finding is in step with the comparatively
small magnitudes of $\Delta s$ obtained in the most well-supported theoretical
calculations, as well as the minimal strangeness permitted by global analyses of high
energy data.


%
%
While slight ambiguity remains as to whether $\Delta s\neq0$, significant improvements
in the last decade on the dual fronts of experimental measurement and theory
have strongly excluded the large magnitudes $\Delta s\sim-0.2$ barely allowed at the
lower reaches of the earliest analyses. As demonstrated above and summarized in
Fig.~\ref{fig:bar}, more recent experiments
suggest $\Delta s\gtrsim-0.1$, whereas theoretical analyses impose 
$\Delta s\gtrsim-0.04$. Most lattice and Bag Model-based calculations especially prefer still
smaller strange helicity magnitudes, $\Delta s\sim-0.02$ --- fully an
order-of-magnitude shallower than the critical value assumed in the
simulations of Ref.~\cite{Melson:2015spa}.
We therefore conclude that, while the nucleon's strange spin is an important
consideration in the microphysics of CCSN simulations, it cannot contribute at the higher level employed
in, {\it e.g.}, Ref.~\cite{Melson:2015spa}; by extension, $\Delta s$ cannot represent the single, decisive effect
generative of CCSN explosions. Rather, for the sake of future numerical simulations of CCSNe, we
advocate the use of more moderate values of $\Delta s$ such as would be supported by the most up-to-date
calculations and measurements. Giving precedence to experimental limits, we suggest $\Delta s=-0.1$ as
a well-motivated figure for which hydrodynamical calculations would be on firm ground. At the same time,
improved experiments may eventually track toward the smaller magnitudes of most theoretical computations,
and we therefore recommend $\Delta s=-0.04$ as a more conservative, auxiliary value in line with such work.
Ultimately, to complement these reduced magnitudes for $\Delta s$, we further urge the exploration of other
potential mechanisms, which might enter at the microscopic level of the relevant nuclear physics or in
the form of heretofore unexplored dynamical effects in the macroscopic structure of CCSNe themselves.

%
%
%
We thank Jeremy Holt, Thomas Janka, Keh-Fei Liu, Sanjay Reddy and Xilin Zhang
for helpful exchanges. The work of TJH and GAM was supported by the
U.S.~Department of Energy Office of Science, Office of Basic Energy Sciences
program under Award Number DE-FG02-97ER-41014. The work of MA was supported
under NSF Grant No.~1205686.
%

%
\end{document}